%% file: CameraReady.tex
\documentclass[10pt,twocolumn,letterpaper]{article}

\usepackage{cvpr}              

\usepackage[accsupp]{axessibility}
\usepackage{graphicx}
\usepackage{amsmath}
\usepackage{amssymb}
\usepackage{booktabs}
\usepackage{bm}
\usepackage{xcolor}
\usepackage{multirow}
\usepackage{kotex}
\usepackage{makecell}

\newcommand{\E}{\operatorname{\mathbb E}}
\newcommand{\R}{\operatorname{\mathbb R}}

\newcommand{\add}[1]{\textcolor{black}{#1\xspace}}

\usepackage[pagebackref,breaklinks,colorlinks]{hyperref}

\usepackage[capitalize]{cleveref}
\crefname{section}{Sec.}{Secs.}
\Crefname{section}{Section}{Sections}
\Crefname{table}{Table}{Tables}
\crefname{table}{Tab.}{Tabs.}

\begin{document}

\title{Joint Global and Local Hierarchical Priors for Learned Image Compression}

\author{
Jun-Hyuk Kim$^1$\thanks{Work done while doing an internship at NAVER AI Lab.} \quad Byeongho Heo$^2$ \quad Jong-Seok Lee$^1$\\
$^1$School of Integrated Technology, Yonsei University \quad $^2$NAVER AI Lab\\
{\tt\small \{junhyuk.kim, jong-seok.lee\}@yonsei.ac.kr} \quad {\tt\small bh.heo@navercorp.com}
}
\maketitle

\begin{abstract}
Recently, learned image compression methods have outperformed traditional hand-crafted ones including BPG. One of the keys to this success is learned entropy models that estimate the probability distribution of the quantized latent representation. Like other vision tasks, most recent learned entropy models are based on convolutional neural networks (CNNs). However, CNNs have a limitation in modeling long-range dependencies due to their nature of local connectivity, which can be a significant bottleneck in image compression where reducing spatial redundancy is a key point. To overcome this issue, we propose a novel entropy model called Information Transformer (Informer) that exploits both global and local information in a content-dependent manner using an attention mechanism. Our experiments show that Informer improves rate--distortion performance over the state-of-the-art methods on the Kodak and Tecnick datasets without the quadratic computational complexity problem. Our source code is available at \url{https://github.com/naver-ai/informer}.
\end{abstract}

\input{1.Introduction}
\input{2.Related_work} 
\input{3.Method}

\input{4.Experiments}

\input{5.Conclusion}

\section*{Acknowledgement}
All experiments were conducted on the NAVER Smart Machine Learning (NSML) platform~\cite{kim2018nsml}.
This work was supported by the NRF grant funded by the Korea government (MSIT) (2021R1A2C2011474).

{\small
\bibliographystyle{ieee_fullname}
\bibliography{egbib}
}

\end{document}

%% file: 1.Introduction.tex
\section{Introduction}
\label{sec:intro}

\begin{figure}[t]
  \centering
  \includegraphics[width=\linewidth]{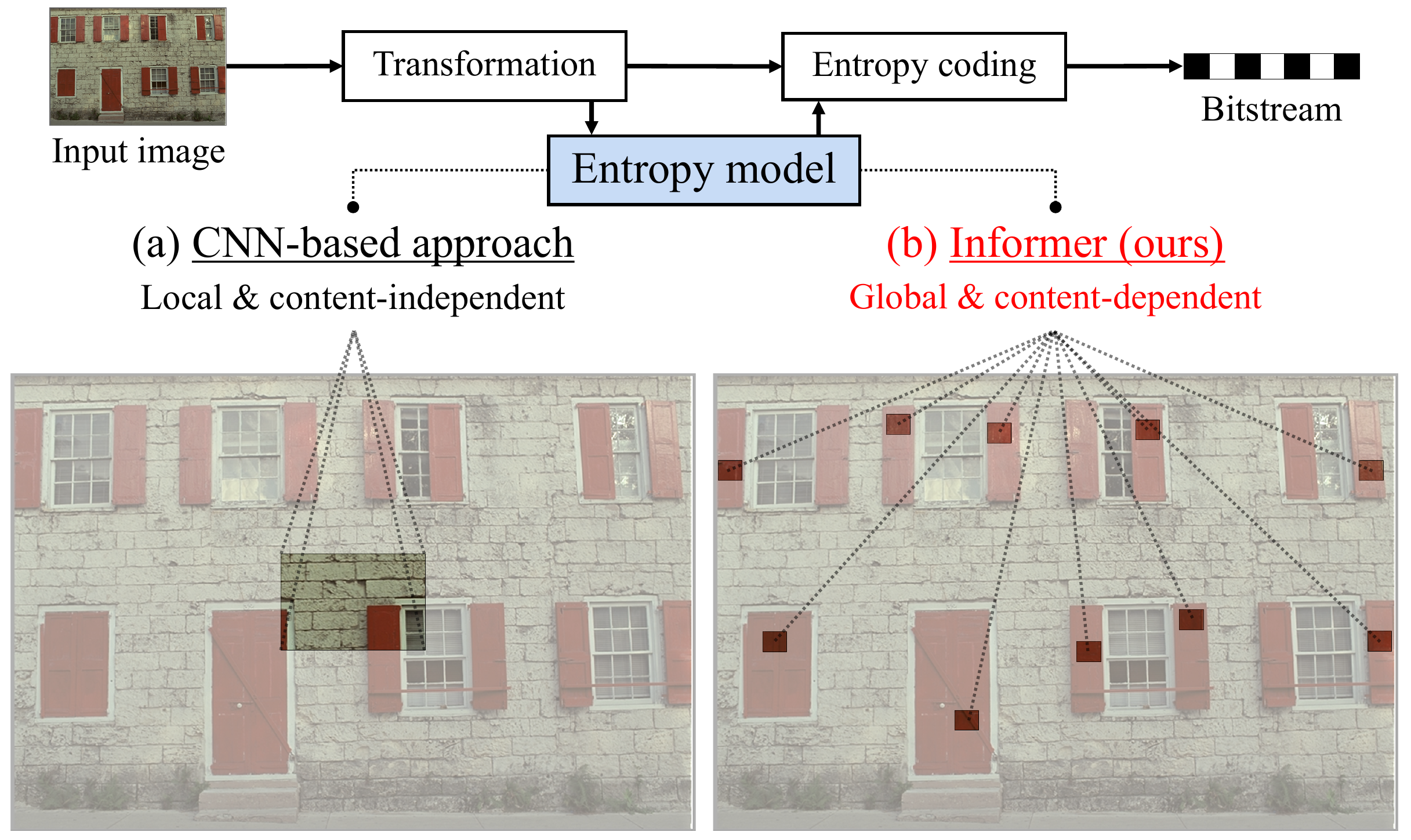}
  \caption{Overview of the proposed method. Our Informer is a learned entropy model capturing global dependencies in a content-dependent manner using the attention mechanism~\cite{vaswani2017attention}.}
  \label{fig:teaser}
  \vspace{-0.7em}
\end{figure}

More than one trillion photos are taken every year and the number is increasing~\cite{blog}, leading to an ever-increasing demand for improved compression efficiency. 
Recently, advances in deep learning have led to significant progress in learned image compression~\cite{toderici2015variable,toderici2017full,johnston2018improved,balle2017end,theis2017lossy,balle2018variational,minnen2018joint,lee2019context,cheng2020learned,qian2021global}.
Generally, learned image compression follows a transform coding framework~\cite{goyal2001theoretical} consisting of transformation and entropy coding (see the top of \cref{fig:teaser}).
In this framework, an image is first transformed into a quantized latent representation that enables more effective compression than the original image.
Then, the quantized latent representation is encoded to a bitstream by a standard entropy coding algorithm (e.g., arithmetic coding~\cite{rissanen1981universal}).
An entropy model, i.e., a prior probability model on the quantized latent representation, is required for the entropy coding algorithm.
Deep neural networks are employed for the transformation and entropy model in this framework~\cite{balle2017end,theis2017lossy,balle2018variational,minnen2018joint,lee2019context,cheng2020learned,qian2021global}, where both are learned in an end-to-end manner to fully utilize the strong capability of deep neural networks~\cite{hornik1989multilayer}.

Since the length of the bitstream relies on the entropy model, designing an accurate entropy model is important for compression efficiency, which is our main focus in this paper.
The goal of entropy models is to estimate a joint probability distribution over the elements of the quantized latent representation.
A simple way to do this is to assume complete independence among the elements~\cite{balle2017end}.
However, this approach yields limited compression efficiency \add{since the assumption does not hold in most practical cases~\cite{balle2018variational}.}
\add{
Thus, how to model the remaining dependencies has been an important issue in learned image compression~\cite{balle2018variational,minnen2018joint,lee2019context,qian2021global}.
}
It is popular to extract additional features, called ``hyperprior'' or ``hierarchical prior'', capturing the dependencies from the latent representation using convolutional neural networks (CNNs).
This approach has contributed to learning accurate entropy models, making learned image compression methods outperform hand-crafted image codes such as BPG~\cite{bpg}.

However, despite the significant progress, CNN-based entropy models still have limitations in capturing the dependencies due to the nature of CNNs.
First, existing entropy models do not make full use of global information due to the local receptive field of CNNs.
This issue can be critical in modeling long-range dependencies.
For example, in the case of \cref{fig:teaser}, the CNN-based approach cannot \add{fully capture the dependencies among} the red windows that repeatedly appear across the whole image due to the localized receptive field.
Second, the receptive fields of previous entropy models cannot exclude nearby elements with different contents due to the content-independent property of convolution operations~\cite{naseer2021intriguing}.
In other words, no matter how different the contents of two elements are, they are processed within the same receptive field if they are located nearby. 
In \cref{fig:teaser}, although the red window and the bricks have quite different contents, both are used simultaneously in the process of capturing dependencies.

To overcome these limitations, we propose a novel entropy model, called Information Transformer (Informer), that captures both global and local dependencies in a content-dependent manner using the attention mechanism of Transformer~\cite{vaswani2017attention} (\cref{fig:teaser}).
In contrast to convolution operations, the attention mechanism has known to be effective in modeling long-range dependencies in a content-dependent manner~\cite{naseer2021intriguing}.
Based on the joint autoregressive and hierarchical priors~\cite{minnen2018joint}, which is the basis of the latest entropy models~\cite{cheng2020learned,qian2021global}, we introduce two novel hyperpriors, i.e., a global hyperprior and a local hyperprior.
To model global dependencies of the quantized latent representation, our Informer first extracts a global hyperprior consisting of different vectors that attend to different areas of an image by using the cross-attention mechanism~\cite{chen2021crossvit,bai2021visual,ma2021luna}.
Furthermore, our Informer extracts a local hyperprior specialized for local information by using 1$\times$1 convolutional layers.
Our local hyperprior prevents our global hyperprior from utilizing only local information and thus allows our Informer to consider global and local information effectively.

Compared to the baseline entropy model~\cite{minnen2018joint}, Informer improves the rate--distortion performance of learned image compression methods on the popular Kodak~\cite{kodak} and Tecnick~\cite{tecnick} datasets.
In addition, Informer achieves better performance than the recently proposed global reference model~\cite{qian2021global} aiming at capturing global dependencies; Informer not only yields higher rate--distortion performance but also avoids the quadratic computational complexity problem of the global reference model.
Our main contributions can be summarized as follows:
\begin{itemize}
    \item We propose joint global and local hyperpriors that effectively model two different types of dependencies between the elements of the quantized latent representation using the attention mechanism. 
    \item We demonstrate that our Informer with the joint global and local hyperpriors improves rate--distortion performance of learned image compression while addressing the quadratic computational complexity problem. 
\end{itemize}

\begin{figure*}[t]
  \centering
  \begin{subfigure}[t]{0.18\linewidth}
    \includegraphics[width=\linewidth]{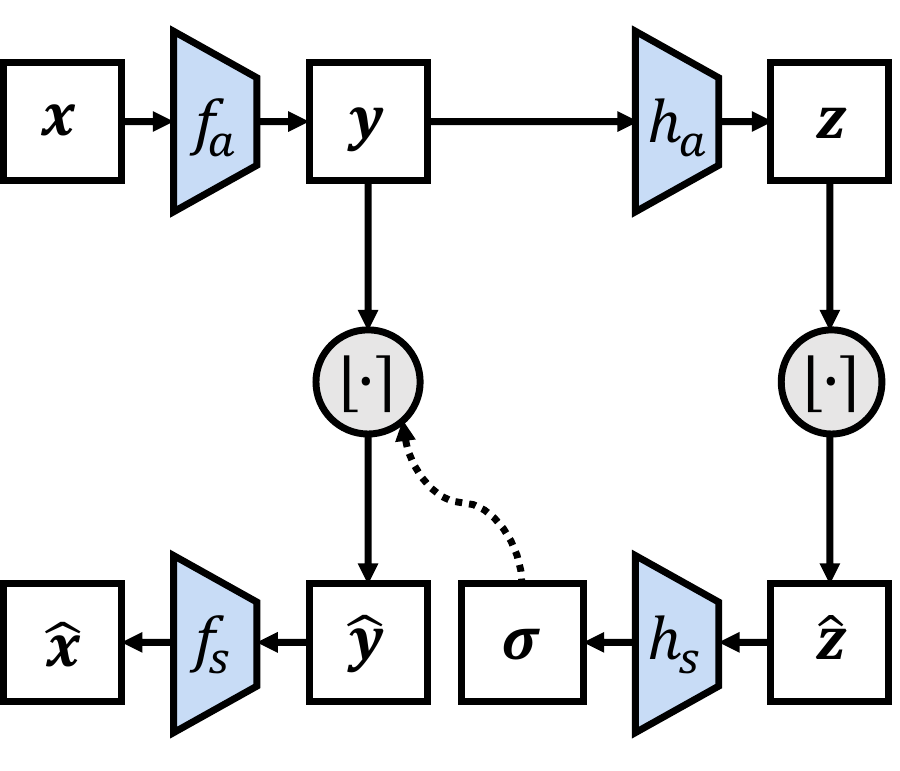}
    \caption{Hyperprior~\cite{balle2018variational}}
    \label{fig:diagram_1}
  \end{subfigure}
  \hfill
  \begin{subfigure}[t]{0.21\linewidth}
    \includegraphics[width=\linewidth]{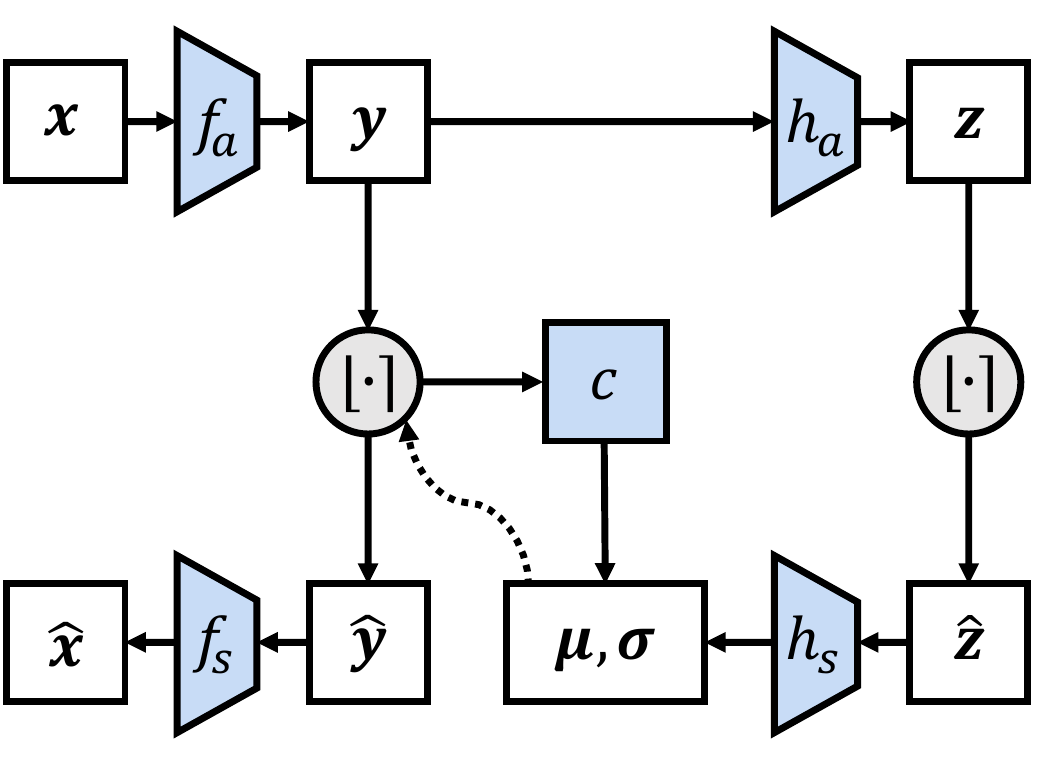}
    \caption{Context and hyperprior~\cite{minnen2018joint}}
    \label{fig:diagram_2}
  \end{subfigure}
  \hfill
  \begin{subfigure}[t]{0.22\linewidth}
    \includegraphics[width=\linewidth]{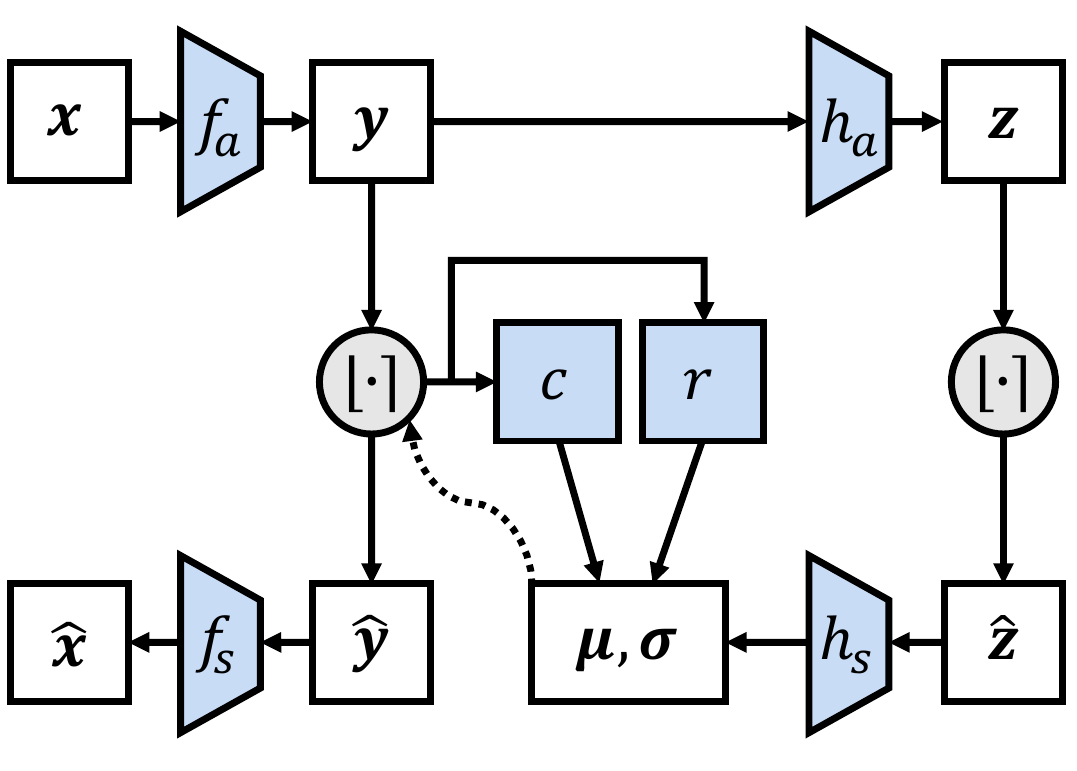}
    \caption{(b)+global reference~\cite{qian2021global}}
    \label{fig:diagram_3}
  \end{subfigure}
  \hfill
  \begin{subfigure}[t]{0.28\linewidth}
    \includegraphics[width=\linewidth]{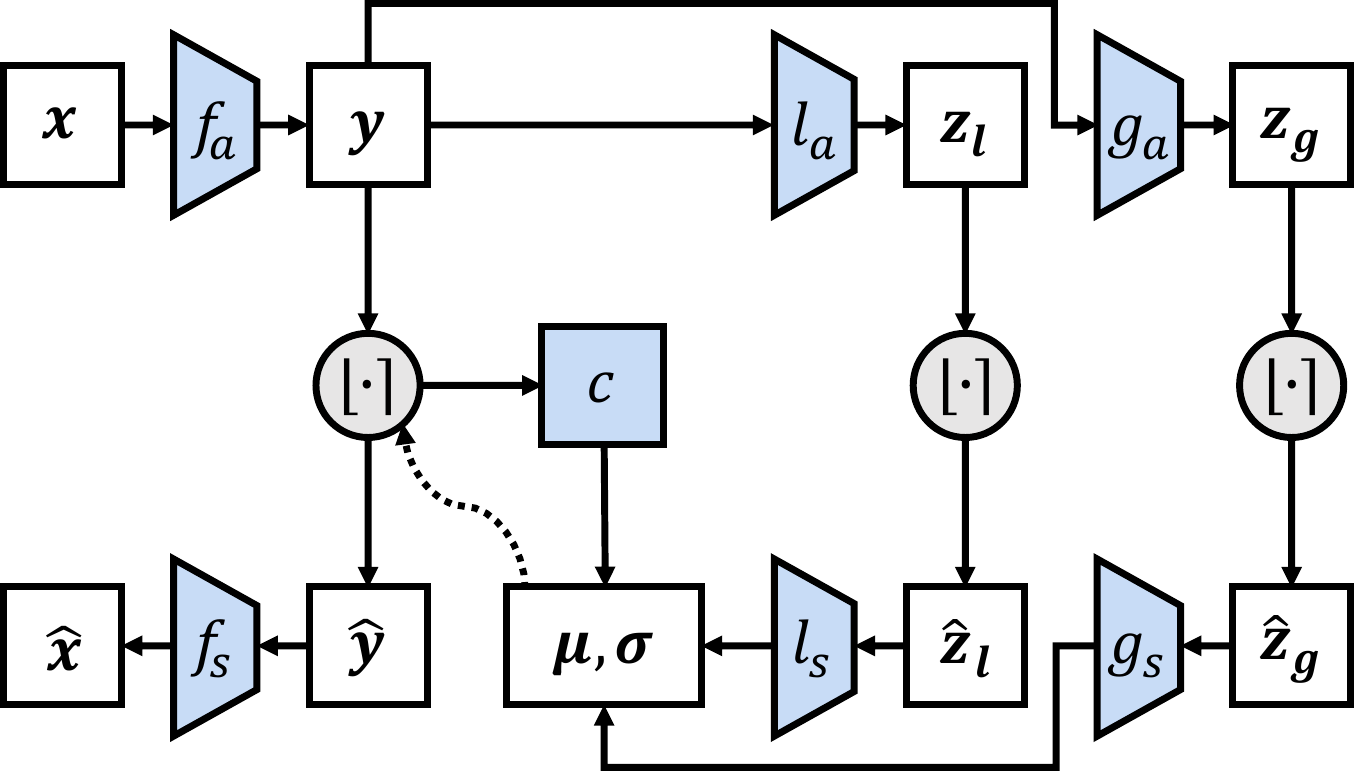}
    \caption{Informer (ours)}
    \label{fig:diagram_4}
  \end{subfigure}
\caption{Operational diagrams of learned image compression methods using different entropy models.
The white blocks are data tensors, the blue blocks represent learned models, and the gray circles mean quantization operations.}
\vspace{-0.7em}
\label{fig:diagram}
\end{figure*}

%% file: 2.Related_work.tex
\section{Related work}
\label{sec:related_work}
\paragraph{Learned nonlinear transforms.}
One of the keys to the success of learned image compression is that deep neural networks effectively model nonlinear transforms suitable for image compression, while traditional image codecs mostly assume linear transforms due to the difficulty of hand-engineering nonlinear transforms for high-dimensional data like images~\cite{balle2020nonlinear}.
Since Ball{\'e}~\etal~\cite{balle2015density} proposed the generalized divisive normalization (GDN) layer that is effective for modeling nonlinear transforms, CNNs with the GDN layers have been widely used in later methods~\cite{balle2017end,balle2018variational,minnen2018joint,lee2019context,kim2020efficient}.
Recently, some learned nonlinear transforms have been proposed using deep residual networks with small kernels (i.e., 3$\times$3)~\cite{cheng2019deep}, an attention module~\cite{cheng2020learned}, invertible neural networks~\cite{xie2021enhanced}, and an attentional multi-scale back-projection module~\cite{gao2021neural}.

\vspace{-1em}
\paragraph{Attention mechanisms.}
The attention mechanism~\cite{vaswani2017attention} is one of the most successful methods to handle global information in deep neural networks. 
It demonstrates notable performance in the language domain through the Transformer architecture~\cite{vaswani2017attention}.
Some researches~\cite{wang2018nonlocal,wang2020axial,kim2020mamnet} tried to utilize the powerful performance of the attention mechanism in the computer vision domain.
Recently, Vision Transformer~\cite{dosovitskiy2020image} achieves state-of-the-art accuracy on image classification tasks.
Many studies have been conducted to use and improve Vision Transformer in diverse vision tasks
such as object detection~\cite{carion2020end,heo2021rethinking,liu2021swin}, semantic segmentation~\cite{zheng2021rethinking,liu2021swin}, and image quality assessment~\cite{cheon2021perceptual}.
Since Transformer has a strong ability to model long-range dependencies regardless of their distance in the pixel domain~\cite{raghu2021vision}, which existing learned entropy models do not have, we propose a novel Transformer-based learned entropy model.

%% file: 3.Method.tex
\section{Joint global and local hyperpriors}
\label{sec:method}
\subsection{Learned image compression}
Given the input image $\bm x$, most learned image compression models~\cite{balle2018variational,minnen2018joint,cheng2020learned} aim to jointly minimize the expected length of the bitstream (i.e., rate) and the expected distortion of the decoded image with respect to $\bm x$:
\begin{equation}
\small
\underbrace{\E_{\bm{x} \sim p_{\bm x}}\bigl[-\log_2 p_{\bm{\hat y}}(\lfloor{f_a(\bm x)}\rceil)\bigr]}_{\text{rate}} + \lambda \cdot 
\underbrace{\E_{\bm x \sim p_{\bm x}}\bigl[d(\bm x, f_s(\lfloor{f_a(\bm x)}\rceil))\bigr]}_{\text{distortion}}.
\label{eq:rate--distortion}
\end{equation}
$\lambda$ is the Lagrange multiplier controlling the trade-off between rate and distortion. $f_a(\cdot)$, $\left\lfloor{\cdot}\right\rceil$, and $f_s(\cdot)$ represent an encoder, a quantizer, and a decoder, respectively. $\bm{\hat y}$ is the quantized latent representation, i.e., $\bm{\hat y}=\lfloor{f_a(\bm x)}\rceil$.
$p_{\bm x}$ is the distribution of training images, and $p_{\bm{\hat y}}$ is the learned entropy model.
When an entropy coding proceeds under the learned entropy model $p_{\bm{\hat y}}$, the smallest rate is the cross-entropy between the actual probability distribution of the quantized latent representation and the learned entropy model $p_{\bm{\hat y}}$.
Thus, the cross-entropy is used for the rate term.
$d(\cdot,\cdot)$ in the distortion term is usually defined by traditional image distortion metrics such as mean squared error (MSE) or multi-scale structural similarity (MS-SSIM). 

To enable gradient-based end-to-end training, studies have been conducted to deal with the non-differentiable quantization operation~\cite{balle2017end,theis2017lossy,agustsson2020universally}. 
The most widely used method is to approximate quantization using additive uniform noise~\cite{balle2017end}, which we adopt in this paper.

\subsection{Learned entropy models}
The entropy models seek to estimate a joint probability distribution over the elements of the quantized latent representation $\bm{\hat y}$.
Note that the rate term in \cref{eq:rate--distortion} is minimized when the learned entropy model $p_{\bm{\hat y}}$ perfectly matches the actual probability distribution.
A simple approach to model the distribution of $\bm{\hat y}$ is to assume that all elements are statistically independent and to learn a fixed entropy model, i.e., fully factorized model~\cite{balle2017end,theis2017lossy}.
Despite its simplicity, this approach does not model the remaining dependencies in $\bm {\hat y}$, and thus cannot achieve optimal performance~\cite{balle2018variational}.

To address this limitation, advanced methods~\cite{balle2018variational,minnen2018joint,lee2019context} propose conditional entropy models where the elements are assumed to follow conditionally independent parametric probability models, and the distribution parameters are adapted by utilizing the remaining dependencies.
They can be divided into two directions: 1) what parametric models to be used~\cite{balle2018variational,minnen2018joint,cheng2020learned,cui2021asymmetric} and 2) how to model dependencies~\cite{balle2018variational,minnen2018joint,lee2019context,qian2021global}.
The former direction includes zero-mean Gaussian~\cite{balle2018variational}, Gaussian~\cite{minnen2018joint}, Gaussian mixture~\cite{cheng2020learned}, and asymmetric Gaussian~\cite{cui2021asymmetric}.
Among them, we employ the widely used one, i.e., Gaussian~\cite{minnen2018joint}.
Specifically, we use a Gaussian distribution convolved with a unit uniform distribution following the previous works~\cite{balle2018variational,minnen2018joint}: 
\begin{equation}
p_{\bm{\hat y}}(\bm{\hat y}) = \prod_i \Bigl( \mathcal N\bigl(\mu_i, \sigma_i^2\bigr) \ast \mathcal U\bigl(-\tfrac 1 2, \tfrac 1 2\bigr) \Bigr)(\hat y_i),
\label{eq:entropy-model}
\end{equation}
where $\mu_i$ and $\sigma_i$ are the mean and scale parameters \add{of the Gaussian distribution for} each element $\hat y_i$, respectively.

The main focus of this study is on the latter direction, i.e., accurate modeling of dependencies.
Existing methods model local dependencies in two different ways.
First, Ball{\'e}~\etal~\cite{balle2018variational} capture the local dependencies by extracting side information that is encoded additionally, which is called a hyperprior.
The operational diagram of this approach is explained in \cref{fig:diagram_1}.
The hyperprior model ($h_a$ and $h_s$) extracts and utilizes the hyperprior $\bm {\hat z}$ for predicting the distribution parameter $\bm \sigma$.
Since additional information is encoded, the rate term in \cref{eq:rate--distortion} is extended as follows:  
\begin{equation}
\E_{\bm x \sim p_{\bm x}}\bigl[-\log_2 p_{\bm{\hat y}}(\bm{\hat y}) - \log_2 p_{\bm{\hat z}}(\bm{\hat z})\bigr],
\label{eq:hyperprior_rate}
\end{equation}
\add{
where the learned entropy model $p_{\bm{\hat z}}$ is designed using the non-parametric fully factorized entropy model~\cite{balle2017end}.
}

Another approach of modeling the local dependencies is to utilize previously decoded adjacent elements (i.e., a context prior~\cite{minnen2018joint,lee2019context}).
While the hyperprior requires additional bits, the context prior is bit-free.
Since Minnen~\etal~\cite{minnen2018joint} and Lee~\etal~\cite{lee2019context} demonstrated that the two kinds of priors are complementary, they have been typically used jointly in literature~\cite{cheng2020learned,cui2021asymmetric,gao2021neural} (\cref{fig:diagram_2}).
The outputs of the context model $c$ and the hyperprior model $h_a$ and $h_s$ are used together for predicting the distribution parameters $\bm \mu$ and $\bm \sigma$.

While the above approaches focus on modeling the local dependencies, Qian~\etal~\cite{qian2021global} propose a global reference model that captures long-range dependencies. 
This utilizes the most relevant previously decoded element for estimating distribution parameters of the current element.
As shown in \cref{fig:diagram_3}, Qian~\etal~\cite{qian2021global} use the global reference model $r$ combined with the joint context and hyperprior model~\cite{minnen2018joint}.

\vspace{-1em}
\paragraph{Motivation of Informer.}
Although utilizing the global dependencies is an innovative direction, the global reference model~\cite{qian2021global} does not fully exploit global information because only a single element among previously decoded ones is used.
To improve the utilization of global information, we introduce a global hyperprior $\bm{\hat z}_g$ \add{extracted from all elements of the latent representation $\bm y$ as shown in \cref{fig:diagram_4}.
}

In addition, the global reference model~\cite{qian2021global} has an issue that the computational complexity increases quadratically with respect to the given image size.
This is because in its self-attention-like mechanism, the most similar element to the current one is searched among all the previously decoded elements \add{and this process is repeated for all elements.}
In order to avoid such an issue, in our global hyperprior modeling, we utilize \add{a cross-attention mechanism} with a fixed number of query regardless of the image size.

\subsection{Decomposition of hyperpriors}
\label{sec:informer}
\begin{figure}[t]
  \centering
  \includegraphics[width=\linewidth]{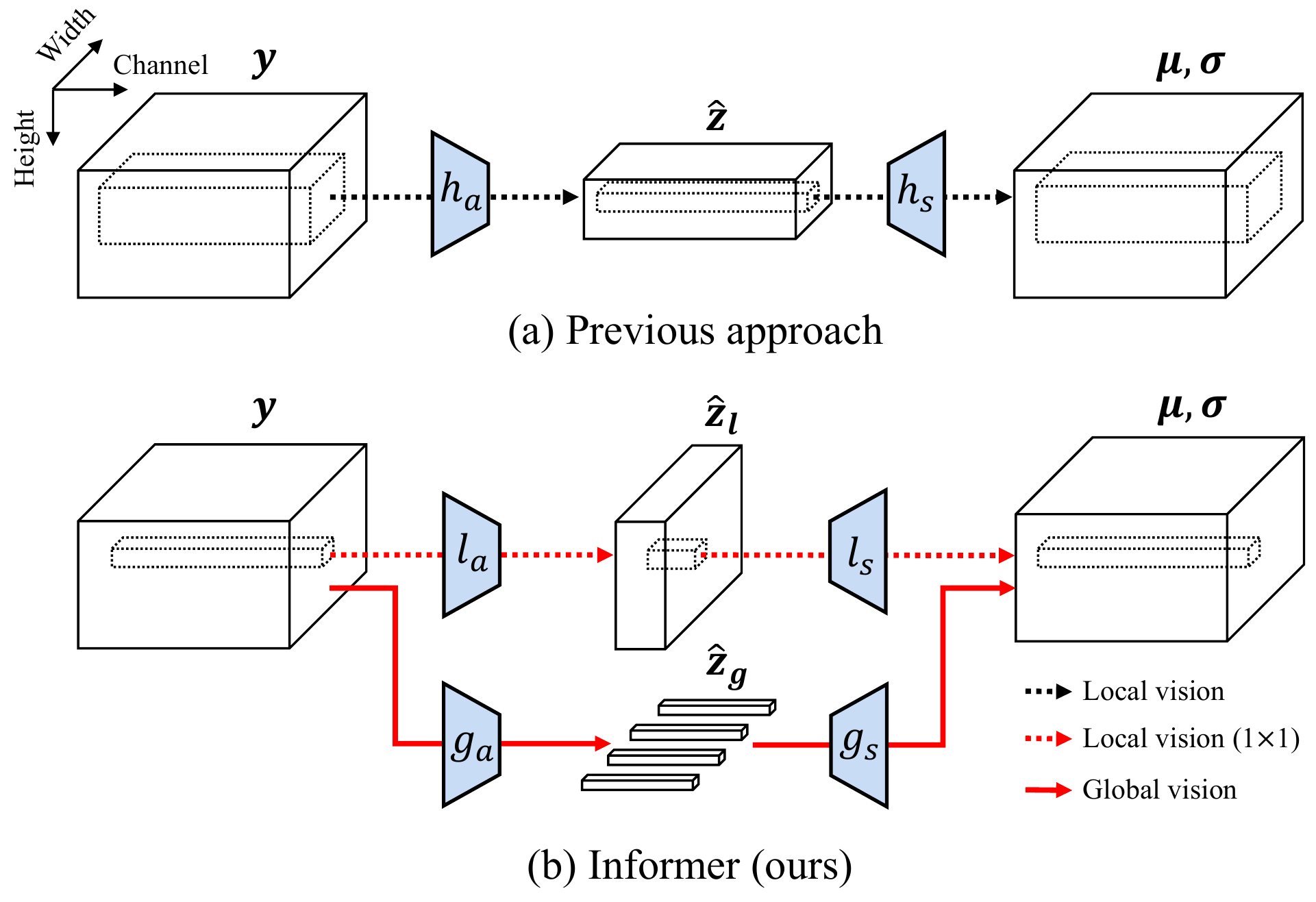}
  \caption{Schematic illustration of a typical hyperprior and the proposed hyperpriors. In contrast to the hyperprior based on spatial dimension reduction~\cite{balle2018variational,minnen2018joint}, our Informer utilizes a global hyperprior $\bm{\hat{z}}_g$ extracted by an attention mechanism and a local hyperprior $\bm{\hat{z}}_l$ specialized for spatial information.
  Quantization operations after $h_a$, $l_a$, and $g_a$ are omitted for simplicity. 
  }
  \label{fig:concept}
  \vspace{-0.7em}
\end{figure}

\add{
As shown in \cref{fig:diagram_4}, extending the context and hyperprior~\cite{minnen2018joint}, our entropy model, Informer, decomposes the hyperprior into two novel hyperpriors: a global hyperprior $\bm{\hat z}_g$ and a local hyperprior $\bm{\hat z}_l$.
}
\cref{fig:concept} illustrates a high-level overview of our hyperpriors in
comparison with the previous approach~\cite{balle2018variational,minnen2018joint}. 
The hyperprior $\bm{\hat z}$ in the previous approach reduces the spatial resolution of the latent representation $\bm y$ while retaining the number of channels.
Due to the localized operations in CNNs, it uses limited local information containing only spatially adjacent elements.

\add{
On the other hand, our global hyperprior $\bm{\hat z}_g$ consists of vectors having no spatial information and is not limited to the local area. Thus, it can handle the whole image area when modeling dependencies.
Specifically, the global dependencies are modeled in a content-dependent manner by using the attention mechanism~\cite{vaswani2017attention}.
In addition, the local hyperprior $\bm{\hat z}_l$ models inter-channel dependencies in each spatial location to complement the lack of spatial components in $\bm{\hat z}_g$. 
It maintains the spatial resolution of the latent representation $\bm y$ while reducing the number of channels.
In summary, the proposed two types of hyperpriors are extracted in parallel capturing dependencies of the latent representation $\bm y$ by effectively complementing each other. 
} 

To extract and utilize the hyperpriors, we build \textit{Global Hyperprior Model} (i.e., \textit{Global Hyper Encoder} $g_a$ and \textit{Global Hyper Decoder} $g_s$) and \textit{Local Hyperprior Model} (i.e., \textit{Local Hyper Encoder} $l_a$ and \textit{Local Hyper Decoder} $l_s$).
\add{
All trainable models (the blue blocks in \cref{fig:diagram_4}) are learned using \cref{eq:rate--distortion} with extending the rate term in order to consider our hyperpriors $\bm{\hat z}_l$ and $\bm{\hat z}_g$:
\begin{multline}
\E_{\bm x \sim p_{\bm x}}\bigl[-\log_2 p_{\bm{\hat y}}(\bm{\hat y}) - \log_2 p_{\bm{\hat z}_l}(\bm{\hat z}_l) - \log_2 p_{\bm{\hat z}_g}(\bm{\hat z}_g)\bigr],
\label{eq:extended_rate}
\end{multline}
where the learned entropy models $p_{\bm{\hat z}_l}$ and $p_{\bm{\hat z}_g}$ are designed using the non-parametric fully factorized entropy model~\cite{balle2017end}.
}


\begin{figure*}[t]
  \centering
  \includegraphics[width=\linewidth]{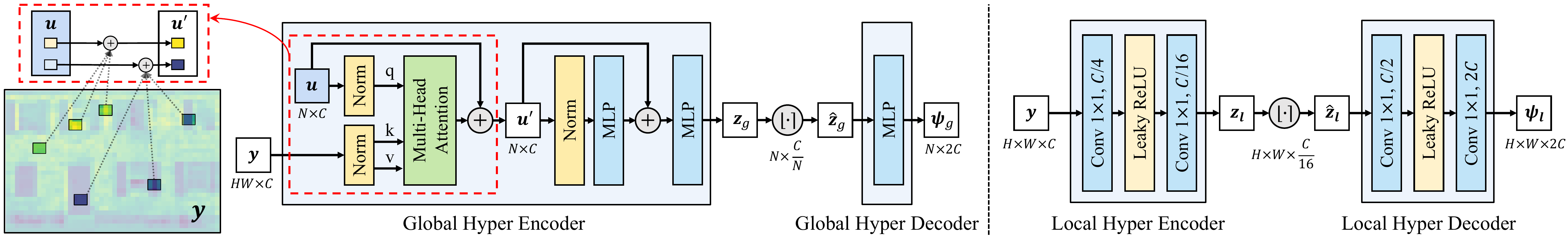}
  \caption{Structure of \textit{Global Hyperprior Model} and \textit{Local Hyperprior Model} in our Informer. The left part illustrates a multi-head attention-based \textit{Global Hyperprior Model}. It utilizes global tokens $\bm u\in\R^{N\times C}$ to extract fixed-size (i.e., $N \times C$) global information from images having various sizes. In the dotted box, we visualize the process of the attention mechanism for the case of $N=2$, where the channel dimension is omitted for simplicity. The global tokens $\bm u$, which are image-independent universal features, are converted to image-dependent features $\bm u'$ by attending the content of the input $\bm y$.
  The right part shows a \textit{Local Hyperprior Model} consisting of $1\times1$ convolutional layers with strides of one that preserve spatial information of images.
  \add{For the convolutional layers, the number of kernels are specified.}}
  \label{fig:hyperpriors}
  \vspace{-0.7em}
\end{figure*}

\subsection{Modeling of hyperpriors}
\paragraph{Global hyperprior.}
Given the input latent representation $\bm y\in\R^{H\times W\times C}$, where $H$, $W$, and $C$ are the height, width, and the number of channels, respectively, the process of extracting the global hyperprior $\bm{\hat z}_g$ is illustrated in \textit{Global Hyper Encoder} in \cref{fig:hyperpriors}.
We define fixed-size global tokens $\bm{u}\in\R^{N\times C}$ as query of the multi-head attention layer of \textit{Global Hyper Encoder}.
$N$ is a predefined parameter, which is fixed to eight in our final models. 
Note that the global tokens are learnable parameters that are determined through end-to-end training like other network parameters.
With the multi-head attention block $MHA(q, k, v)$, the MLP block $MLP_{1}(\cdot)$, and the last MLP layer $MLP_{2}(\cdot)$, the formulation of our \textit{Global Hyper Encoder} is as follows:
\begin{equation}
\begin{aligned}
    \bm{u}' &= MHA(\bm{u}, \bm{y}, \bm{y}), \\
    \bm{z}_g &= MLP_2(MLP_1(\bm{u}')),
\end{aligned}
\end{equation}
where the normalization layers~\cite{ba2016layernorm} are omitted for simplicity. 
\add{$MHA(q, k, v)$ models global dependencies.}
$MLP_{1}(\cdot)$ and $MLP_{2}(\cdot)$ extract $\bm z_g\in\R^{N\times\frac{C}{N}}$ for further modeling of inter-channel dependencies.
After the quantization operation $\lfloor\cdot\rceil$, the global hyperprior $\bm{\hat z}_g$ is obtained. 
\textit{Global Hyper Decoder}, which is one linear layer, receives the global hyper prior $\bm{\hat z}_g$ as the input and generates $\bm {\psi}_g\in\R^{N\times2C}$.

\vspace{-1em}
\paragraph{Local hyperprior.}
In order to model inter-channel dependencies at each spatial location, we design \textit{Local Hyperprior Model} by stacking 1$\times$1 convolutional layers with the leaky ReLU activation~\cite{leakyrelu}, which is shown in the right part of \cref{fig:hyperpriors}.
From the input $\bm y$, \textit{Local Hyper Encoder} extracts the local hyperprior $\bm{\hat z}_{l}\in\R^{H\times W\times \frac{C}{16}}$.
\textit{Local Hyper Decoder} utilizes $\bm{\hat z}_l$ and yields its output $\bm \psi_l\in\R^{H\times W\times 2C}$.


\subsection{Prediction of distribution parameters}
\label{sec:overview}
As shown in \cref{fig:diagram_4}, we also employ \textit{Context Model} $c$ that uses the previously decoded elements $\bm {\hat y}_{<i}$. 
We adopt the same structure as in Minnen~\etal~\cite{minnen2018joint}, i.e., one 5$\times$5 masked convolutional layer.
For predicting the distribution parameters, i.e., mean $\bm \mu \in\R^{H\times W\times C}$ and scale $\bm \sigma \in\R^{H\times W\times C}$, we combine the outputs from \textit{Context Model}, \textit{Local Hyperprior Model}, and \textit{Global Hyperprior Model}. 
Since the output of \textit{Global Hyperprior Model} $\bm{\psi}_g\in\R^{N\times2C}$ does not have the spatial dimension, conventional feature combining methods such as concatenation or element-wise addition are inapplicable. 
Therefore, we propose a new \textit{Parameter Model} with a multi-head attention-based combining method, which is shown in \cref{fig:parameter_model}.
The outputs of \textit{Context Model} and \textit{Global Hyperprior Model} ($\bm \phi$ and $\bm{\psi}_g$, respectively) are combined by using the multi-head attention block. 
After the multi-head attention block and following MLP block, the result is concatenated with the output of \textit{Local Hyperprior Model} $\bm{\psi}_l$ and the distribution parameters $\bm \mu$ and $\bm \sigma$ are generated via three 1$\times$1 convolutional layers.
The leaky ReLU activation functions are used after the first two convolutional layers, which are omitted in \cref{fig:parameter_model}. 
The formulation of \textit{Parameter Model} is as follows:
\begin{equation}
\begin{aligned}
    \{\bm \mu, \bm \sigma\} &= pm(\bm{\phi}, \bm{\psi}_g, \bm{\psi}_l) \\
    \text{with } \bm{\phi} &= c(\bm{\hat y}_{<i}),\text{ } 
    \bm{\psi}_g = gh(\bm y), \text{ and }
    \bm{\psi}_l = lh(\bm y),
\end{aligned}
\end{equation}
where $pm(\cdot)$, $c(\cdot)$, $gh(\cdot)$, and $lh(\cdot)$ represent \textit{Parameter Model}, \textit{Context Model}, \textit{Global Hyperprior Model}, and \textit{Local Hyperprior Model}, respectively.

%% file: 4.Experiments.tex
\section{Experiments}
\label{sec:exp}
All experiments are conducted on a PyTorch~\cite{pytorch} based open-source library: CompressAI platform~\cite{begaint2020compressai}, which has recently been introduced for developing and evaluating learning-based image codecs.

\vspace{-1em}
\paragraph{Training.}
Our models are trained with various configurations of the Lagrange multiplier $\lambda$ and the distortion metric $d(\cdot, \cdot)$.
We use MSE and $(1-\text{MS-SSIM})$ for $d(\cdot, \cdot)$, and set $\lambda\in\{0.0018,0.0035,0.0067,0.0130,0.0250,0.0483\}$ for MSE and $\lambda\in\{2.40,4.58,8.73,16.64,31.73,60.50\}$ for MS-SSIM; thus, a total of 12 final models are obtained.
We use the training images of the Triplet dataset, which is a subset of Vimeo-90K~\cite{xue2019video}.
We use a batch size of 8 with 256$\times$256 patches randomly cropped from the training images.
All models are trained using the Adam optimizer~\cite{kingma2014adam} for 250 epochs. 
The learning rate starts at $10^{-4}$ and decreases to one-third at the 150th, 180th, 210th, and 240th epochs. 
For the MS-SSIM-optimized models, we finetune the MSE-optimized models with a learning rate of $10^{-5}$. 

\vspace{-1em}
\paragraph{Evaluation.}
We evaluate our method on the Kodak~\cite{kodak} and Tecnick~\cite{tecnick} datasets, which are commonly used for benchmarking learned image compression methods. 
The Kodak dataset consists of 24 images with a resolution of 768$\times$512 pixels. 
The Tecnick dataset includes 100 images with a resolution of 1200$\times$1200 pixels.
At the encoding stage, we use the asymmetric numeral systems~\cite{duda2013asymmetric} for the entropy coding.
To evaluate rate--distortion performance, we use the bits per pixel (bpp) and either the peak signal-to-noise ratio (PSNR) or MS-SSIM depending on the distortion metric $d(\cdot, \cdot)$.
MS-SSIM is converted into decibels, i.e., $-10\log_{10}(1-\text{MS-SSIM})$~\cite{balle2018variational}.

\begin{figure}[t]
  \centering
  \includegraphics[width=\linewidth]{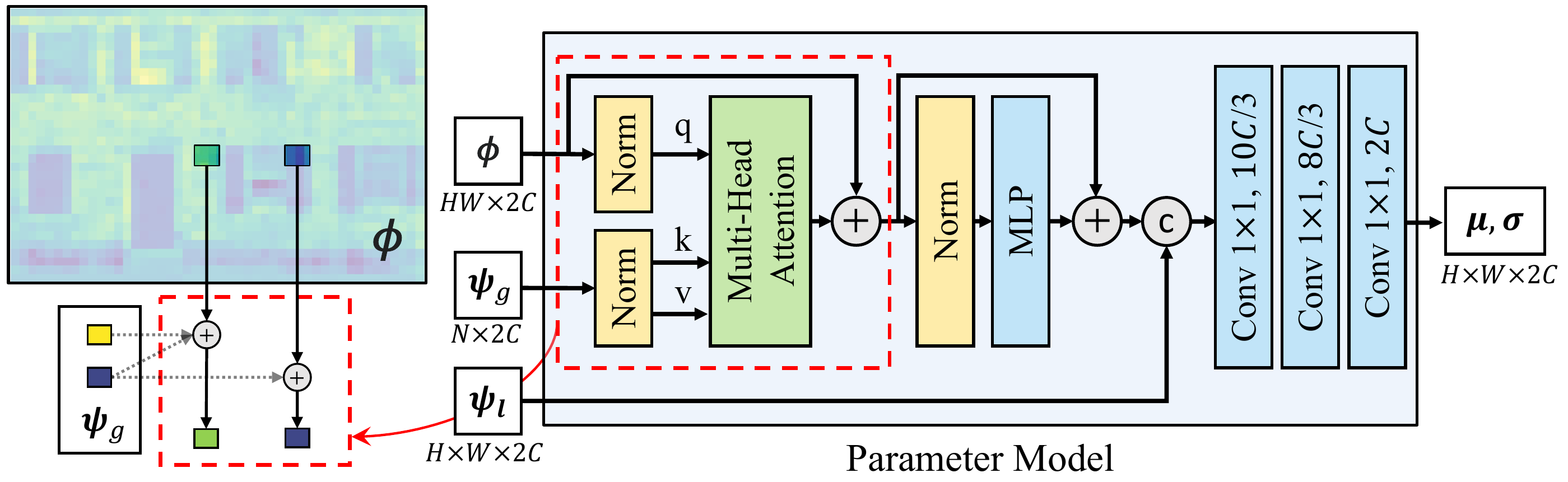}
  \caption{Structure of \textit{Parameter Model}. It combines the outputs of \textit{Context Model}, \textit{Global Hyperprior Model}, and \textit{Local Hyperprior Model} ($\bm \phi$, $\bm{\psi}_g$, and $\bm{\psi}_l$, respectively), and predicts the distribution parameters (mean $\bm \mu$ and scale $\bm \sigma$). 
  \add{
  In the dotted box, we visualize the process of the attention mechanism ($N=2$), where the channel dimension is omitted for simplicity. The local location-specific information $\bm \phi\in\R^{H\times W\times 2C}$ is updated by attending global location-free information $\bm{\psi}_g\in\R^{N\times 2C}$. For the  convolutional layers, the number of kernels are specified.
  }
  }
  \vspace{-0.7em}
  \label{fig:parameter_model}
\end{figure}

\begin{figure*}[t]
  \centering
  \includegraphics[width=\linewidth]{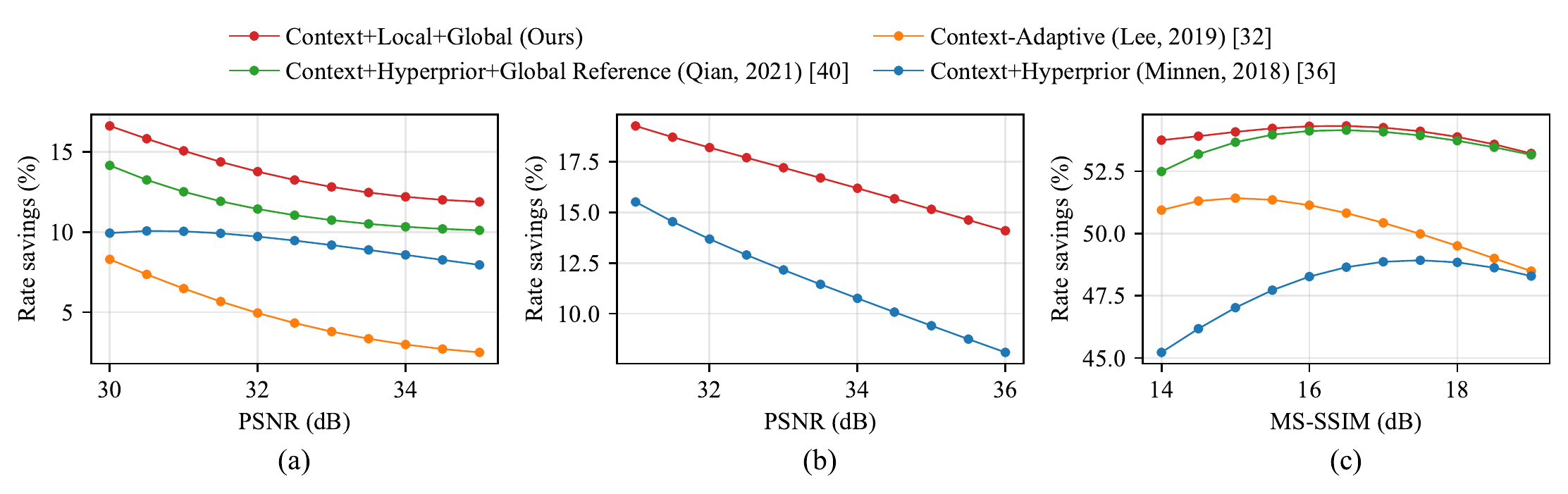}
  \vspace{-2em}
  \caption{Performance of image compression methods using different entropy models. Each curve represents the rate savings ($\%$) relative to BPG~\cite{bpg} at different quality levels. Larger values mean better performance. The results of the MSE-optimized methods are averaged over (a) Kodak~\cite{kodak} and (b) Tecnick~\cite{tecnick}, respectively, and (c) the results of MS-SSIM-optimized methods are averaged on Kodak~\cite{kodak}.
  } 
  \label{fig:rd_performance}
\end{figure*}

\begin{table*}[t]
    \centering
    \scalebox{0.85}{
    \renewcommand{\arraystretch}{1.1}
    \begin{tabular}{lccccccc}
    \toprule
    Method & 320$\times$240 & 480$\times$360 & 640$\times$480 & 768$\times$512 & 1280$\times$720 & 1920$\times$1080 & 4096$\times$2304 \\
    \midrule
    Context+Hyperprior~\cite{minnen2018joint} & 2.73 & 6.14 & 10.91 & 13.96 & 32.73 & 73.64 & 335.13\\
    Context+Hyperprior+Global Reference~\cite{qian2021global} & 9.98 & 22.85 & 41.59 & 54.03 & 138.04 & 366.57 & 3296.99 \\
    Informer (ours) & \textbf{2.69} & \textbf{6.04} & \textbf{10.73} & \textbf{13.73} & \textbf{32.15} & \textbf{72.34} & \textbf{329.17}\\
    \bottomrule
    \end{tabular}
    }
    \caption{
    Comparison of complexity of different entropy models at various image sizes. Each value means GFLOPs of each entropy model when the image having the corresponding resolution is used for entropy modeling. 
    }
    \label{tab:complexity}
\end{table*}

\begin{figure*}[!t]
  \centering
  \includegraphics[width=\linewidth]{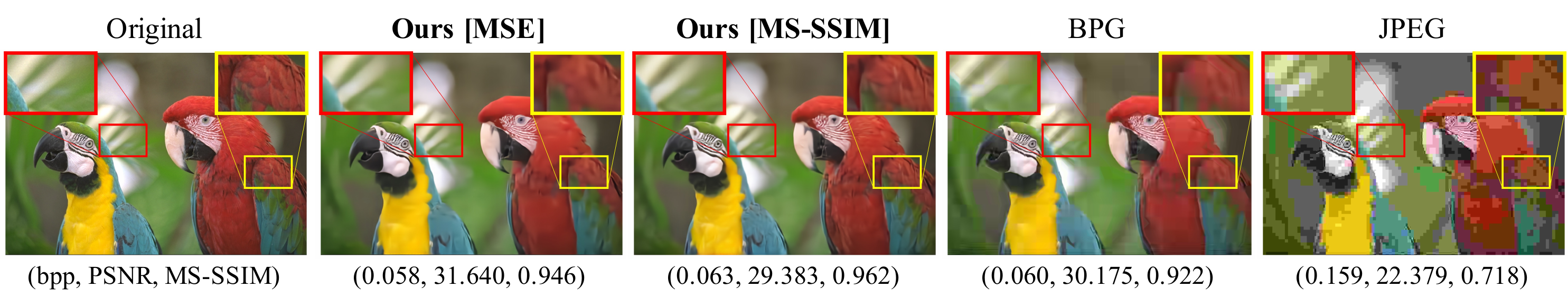}
  \vspace{-2em}
  \caption{Visual comparison of the decoded images by our methods and the other image codecs on ``Kodim23'' from Kodak~\cite{kodak}.}
  \label{fig:visual_results}
  \vspace{-0.7em}
\end{figure*}

\subsection{Performance comparison}
\paragraph{Comparison with other entropy models.}
To show the effectiveness of Informer, we compare its rate--distortion performance with that of the state-of-the-art image compression methods~\cite{minnen2018joint,lee2019context,qian2021global} in which the transformation parts are modeled similar to that of Minnen~\etal~\cite{minnen2018joint} and different entropy models are used.
Each of their encoder and decoder consists of four convolutional layers.
Except for the method of Qian~\etal~\cite{qian2021global}, which uses its own generalized subtractive and divisive normalization (GSDN), all methods use GDN~\cite{balle2015density} as activation functions.
But, for a fair comparison, we use the values of the variant of Qian~\etal~\cite{qian2021global} using GDN, which is reported in their paper. 
Because Qian~\etal~\cite{qian2021global} do not report the performance of the variant optimized with MS-SSIM, we use the values of the MS-SSIM-optimized full models using GSDN in their paper.

\cref{fig:rd_performance} shows the relative rate savings compared to BPG~\cite{bpg} at different quality levels.
Note that these graphs are generalized versions of the popular Bj{\o}ntegaard Delta (BD) chart~\cite{bjontegaard2001calculation} used for evaluation of image compression.
As shown in Figs.~\ref{fig:rd_performance}\textcolor{red}{a} and \textcolor{red}{b}, our MSE-optimized models show significant rate savings over BPG, ranging from 11.84$\%$ to 19.27$\%$.
Compared to the baseline entropy model, i.e., ``Context+Hyperprior (Minnen, 2018)~\cite{minnen2018joint}'', Informer significantly improves rate savings across all PSNR levels on both datasets.
This shows that our global and local hyperpriors improve the accuracy of the learned entropy model.
Even compared to the entropy model aiming at modeling global dependencies, i.e., ``Context+Hyperprior+Global Reference (Qian, 2021)~\cite{qian2021global}'', Informer shows superiority across all PSNR levels on Kodak~\cite{kodak} with performance gaps from 1.74$\%$ up to 2.60 $\%$.
These results demonstrate the effectiveness of Informer's capability to model joint global and local dependencies.
Furthermore, when optimized with MS-SSIM (\cref{fig:rd_performance}\textcolor{red}{c}), Informer outperforms all the other models on Kodak~\cite{kodak}, which achieves rate savings by large margins up to 54.32$\%$.

\vspace{-1em}
\paragraph{Complexity.}
We evaluate the performance of our Informer in terms of computational efficiency. 
For this, we \add{calculate} the number of floating-point operations (FLOPs) required for various image sizes.
\cref{tab:complexity} shows GFLOPs for our Informer and other entropy models~\cite{minnen2018joint,qian2021global}.
Informer shows the best performance across all image sizes.
In particular, Informer does not suffer from the quadratic computational complexity problem that the ``Context+Hyperprior+Global Reference~\cite{qian2021global}'' model has.
In other words, as the image resolution increases, GFLOPs of ``Context+Hyperprior+Global Reference~\cite{qian2021global}'' increases quadratically, being 10 times larger at the resolution of 4096$\times$2304 than at the resolution of 1920$\times$1080, while the complexity of Informer has a near-linear scale.
\add{This is because Informer models global dependencies using a cross-attention mechanism with a fixed number of query.}

\vspace{-1em}
\paragraph{Qualitative performance.}
We compare visual quality of our decoded images with those of other traditional image codecs on the Kodak dataset~\cite{kodak}.
For the image codecs, we employ JPEG~\cite{wallace1992jpeg} and BPG~\cite{bpg}.
For a fair comparison, we encode the images under compression settings for as similar bpp values as possible.
\cref{fig:visual_results} shows that our method produces clear patterns (red boxes) and does not lose details (yellow boxes) compared to the other results.




\begin{figure}[t]
  \centering
    \includegraphics[width=\linewidth]{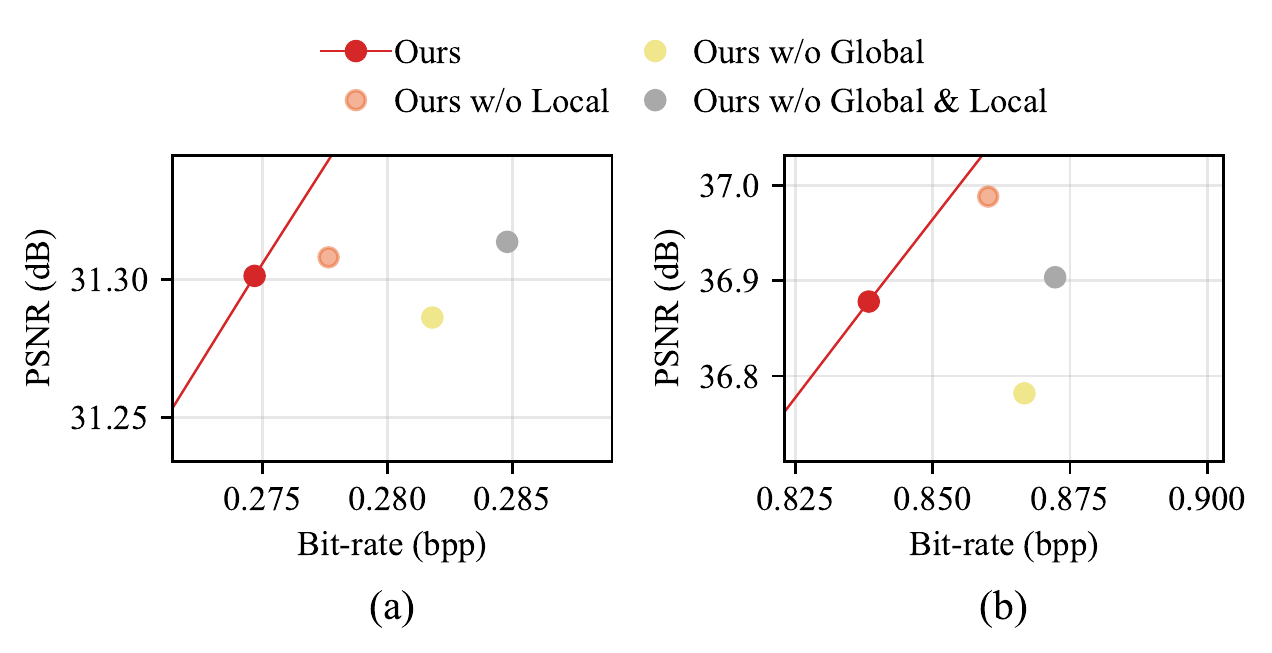}
    \vspace{-2em}
  \caption{
  \add{
  Ablation studies of our global and local hyperpriors in Informer. (a) $\lambda=0.0067$ and (b) $\lambda=0.0483$.
  }
  }
  \label{fig:ablation}
  \vspace{-0.7em}
\end{figure}

\subsection{Model analysis}
We perform further analysis of Informer. 
For this, we utilize additional MSE-optimized models in two different bpp regions using $\lambda=0.0067$ and $\lambda=0.0483$, respectively.

\vspace{-1em}
\paragraph{Ablation study.}
To evaluate the contribution of the proposed hyperpriors in Informer, we conduct ablation studies.
The rate--distortion performance with or without each hyperprior is shown in \cref{fig:ablation}.
It is shown that the original Informer shows the best results regardless of the bpp region.
Introducing our global hyperprior leads to significant improvement compared to the entropy model using only the context prior (i.e., ``Ours w/o Global \& Local''), while our local hyperprior seems to be more helpful when combined with our global hyperprior than when used alone.

\begin{figure}[t]
  \centering
    \includegraphics[width=\linewidth]{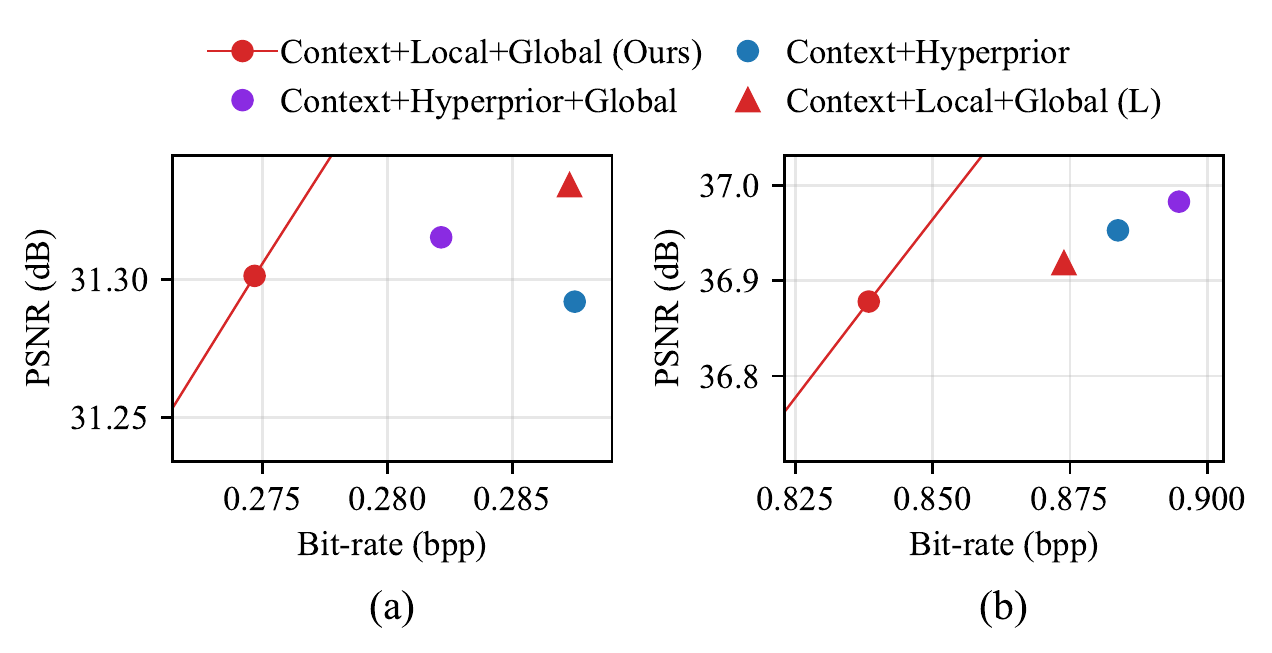}
    \vspace{-2em}
  \caption{
  \add{
  Comparison of different decomposition methods (circles with different colors) and different combining methods (red markers with different shapes). (a) $\lambda=0.0067$ and (b) $\lambda=0.0483$.
  }
  }
  \label{fig:decomposition}
  \vspace{-0.7em}
\end{figure}

\begin{figure}[t]
  \centering
    \includegraphics[width=\linewidth]{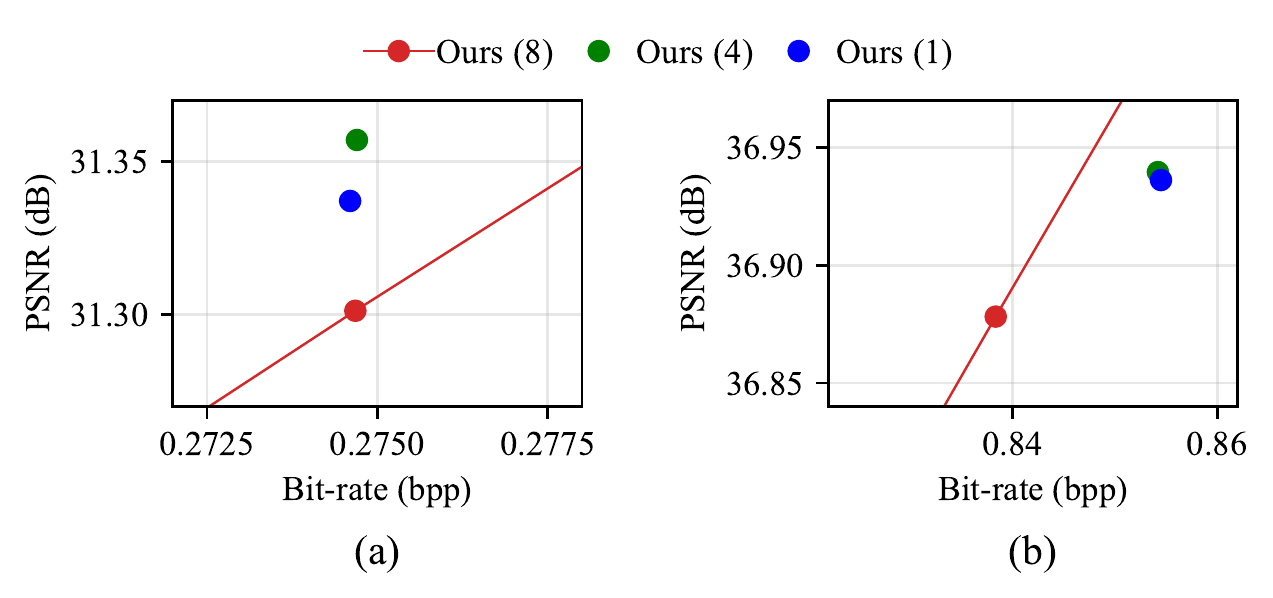}
    \vspace{-2em}
  \caption{Model analysis with respect to the number of global tokens $N$, which is indicated by the number in parentheses. (a) $\lambda=0.0067$ and (b) $\lambda=0.0483$.}
  \label{fig:num_global}
  \vspace{-0.7em}
\end{figure}

\vspace{-1em}
\paragraph{Analysis of decomposition.}
To validate the effectiveness of our approach to decomposing hyperpriors, we compare two different decomposition methods in \cref{fig:decomposition}: Informer (``Context+Global+Local'') and the existing hyperprior with our global hyperprior (``Context+Hyperprior+Global'').
As a reference, we also provide the results of the existing context and hyperprior model~\cite{minnen2018joint} (``Context+Hyperprior''), which uses a single hyperprior without decomposition.
We observe that Informer is better than the others for both bpp regions.
The ``Context+Hyperprior+Global'' method shows even similar or slightly worse rate--distortion performance than the baseline (``Context+Hyperprior'') in the higher bpp region.
We conjecture that this is due to the overlap in the roles of the existing hyperprior and our global hyperprior.
Since the hyperprior-based approach requires additional \add{bit allocation}, redundant roles can be an obstacle for improving performance due to excessive increase of bit usage, which implies that elaborate modeling of hyperpriors is important.

\vspace{-1em}
\paragraph{Analysis of combining methods.}
In \cref{fig:decomposition}, Informer using a variant of \textit{Parameter Model}, denoted as ``Context+Local+Global (L)'', is also evaluated, which combines the inputs to \textit{Parameter Model} in a different way. 
Specifically, the output of \textit{Local Hyperprior Model} $\bm{\psi}_l$ is used as query for the attention layer, and the output of \textit{Context Model} $\bm \phi$ is concatenated after the MLP block.
Our method shows better performance than this variant. 
\add{
In other words, the global hyperprior is more effective when used for updating the context prior than updating the local hyperprior.  
}

\vspace{-1em}
\paragraph{Analysis of attention.}
We examine the effect of the number of global tokens $N$ on rate--distortion performance in \cref{fig:num_global}.
We observe that using four global tokens (``Ours (4)'') yields the best performance at the lower bpp region, while the best performance at the higher bpp region is obtained using eight global tokens (``Ours (8)'').
This is probably due to the different degrees of remaining dependencies between the quantized latent representation $\bm{\hat y}$.
In other words, a large number of global tokens is effective to capture detailed dependencies for the higher bpp region.

\begin{figure}[t]
  \centering
  \includegraphics[width=1.0\linewidth]{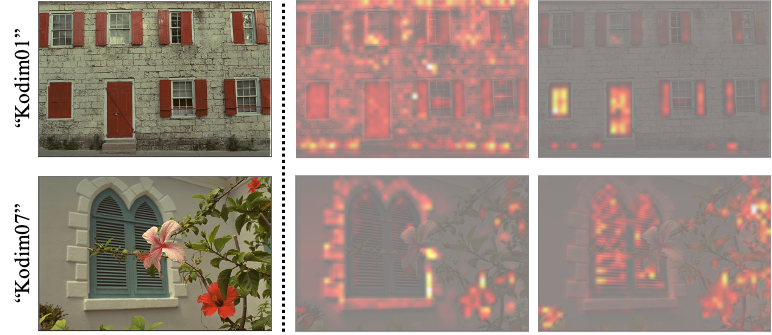}
  \caption{Visualization of the attention map used by the attention mechanism of the proposed \textit{Global Hyper Encoder}. Two sample images from Kodak~\cite{kodak} are used, and two attention maps corresponding to two example global tokens are shown.}
  \label{fig:attention_map}
  \vspace{-0.7em}
\end{figure}

In addition, \cref{fig:attention_map} visualizes attention maps used by the attention mechanism of our \textit{Global Hyper Encoder}, which show where the example global tokens pay attention to for capturing global dependencies. 
The results show that the global tokens successfully capture global information over the whole image area and different global tokens utilize different image regions in a content-dependent manner.
\add{
For example, in the case of ``kodim07'', the global token shown on the left side attends to the bricks located along the edge of the window, while the global token shown on the right side attends to the window. 
}

\subsection{Global hyperprior vs. global context prior}
\begin{figure}[t]
  \centering
  \includegraphics[width=\linewidth]{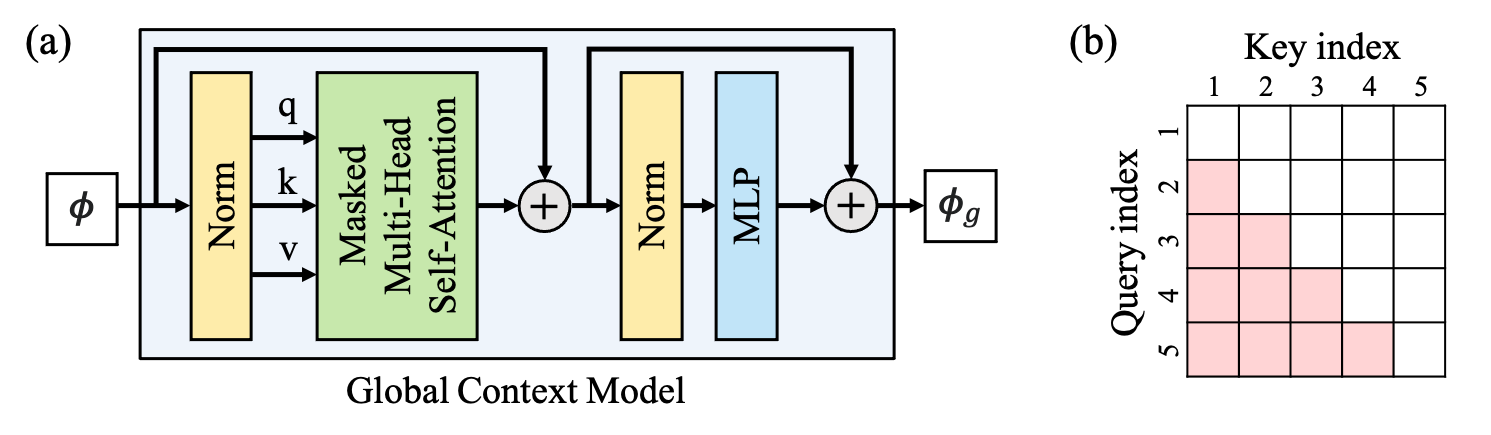}
  \caption{Global context model. (a) Structure and (b) an example of masked self-attention map where the red boxes are activated.}
  \label{fig:global_context_model}
\end{figure}

To examine whether the superior performance of Informer originates from the strong capability of the attention mechanism-based structure or introduction of the joint global and local hyperpriors, we evaluate another entropy model aiming at capturing global dependencies.
Building on the context and hyperprior entropy model~\cite{minnen2018joint} as in Informer, we newly design \textit{Global Context Model} using the masked multi-head self-attention mechanism~\cite{vaswani2017attention}.
As shown in \cref{fig:global_context_model}\textcolor{red}{a}, \textit{Global Context Model} receives the output of \textit{Context Model} $\bm \phi$ and updates it by attending the previously decoded elements. 
Through the process of the attention mechanism, \textit{Global Context Model} considers multiple references, and thus it can be seen as a generalized version of the global reference model~\cite{qian2021global} utilizing only the most relevant reference. 
\cref{tab:compare_global} shows that Informer is superior to the method using \textit{Global Context Model}.
We argue that introducing the self-attention mechanism is not a silver bullet for image compression, but careful design consideration should be accompanied with it.

\begin{table}[t]
    \centering
    \scalebox{0.95}{
    \renewcommand{\arraystretch}{1.2}
    \begin{tabular}{clcc}
    \toprule
    $\lambda$ & Method & Rate (bpp) & PSNR (dB) \\
    \midrule
    \multirow{2}{*}[-2pt]{0.0067} & Informer & \textbf{0.2763} & \textbf{31.3254} \\
     & Global context & 0.2842 & 31.2947 \\
    \midrule
    \multirow{2}{*}[-2pt]{0.0483} & Informer & \textbf{0.8514} & \textbf{36.9739} \\
     & Global context & 0.8718 & 36.9134 \\
    \bottomrule
    \end{tabular}
    }
    \caption{
    Comparison of the different methods of introducing global vision to learned entropy models. The ``Global context'' model introduces global vision to \textit{Context Model} using the masked self-attention mechanism.
    The results are obtained at 160 epochs.
    }
    \label{tab:compare_global}
    \vspace{-0.7em}
\end{table}

%% file: 5.Conclusion.tex
\section{Conclusion}
\label{sec:conclusion}
We proposed a novel learned entropy model (Informer) for learned image compression.
Based on the previous joint autoregressive and hierarchical priors~\cite{minnen2018joint}, Informer introduced two different hyperpriors for modeling remaining dependencies in the quantized latent representation, one for global dependencies and the other for local dependencies.
We showed that Informer outperforms existing entropy models in terms of rate--distortion performance with computational efficiency. 
Beyond the existing CNN-based localized dependency modeling methods, Informer presents a completely new approach that effectively utilizes both global and local information in a content-dependent manner using the attention mechanism.

\section{Limitation}
Informer cannot be parallelized in its decoding process, which is due to the inherent limitation of the autoregressive prior that utilizes only previously decoded elements.
We expect that this problem can be addressed by combining our approach with the channel-wise autoregressive model~\cite{minnen2020channel} or bidirectional context model~\cite{he2021checkerboard}.